\documentclass[10pt,letterpaper]{article}
\usepackage{opex3}

\usepackage{graphicx}
\usepackage{bm}
\usepackage{epsfig}

\def\e{\begin{equation}}
\def\f{\end{equation}}
\def\=#1{\overline{\overline #1}}

\def\-#1{{\bf #1}}
\def\.{\cdot}
\def\l#1{\label{eq:#1}}
\def\r#1{(\ref{eq:#1})}
\def\vec#1{{\bf #1}}

\begin{document}

\title{Spatially dispersive finite-difference time-domain analysis of sub-wavelength imaging by the wire medium slabs}

\author{Yan Zhao, Pavel A. Belov and Yang Hao}

\address{Queen Mary University of London, Mile End Road, London, E1 4NS, United Kingdom}

\email{yan.zhao@elec.qmul.ac.uk} 



\begin{abstract}
In this paper, a spatially dispersive finite-difference time-domain
(FDTD) method to model wire media is developed and validated.
Sub-wavelength imaging properties of the finite wire medium slabs
are examined. It is demonstrated that the slab with its thickness
equal to an integer number of half-wavelengths is capable of
transporting images with sub-wavelength resolution from one
interface of the slab to another. It is also shown that the
operation of such transmission devices is not sensitive to their
transverse dimensions, which can be made even comparable to the
wavelength. In this case, the edge diffractions are negligible and
do not disturb the image formation.
\end{abstract}

\ocis{(120.6640) Superresolution, (110.2990) Image formation
theory, (260.2030) Dispersion.} 

\bibliographystyle{osajnl}

\section{Introduction}
The conventional imaging systems operate only with propagating
harmonics emitted by a source, since the spatial harmonics
carrying sub-wavelength information (evanescent waves) exhibit
exponential decay in free space. This is why the resolution of
common imaging systems is restricted by the diffraction limit: The
details smaller than the wavelength can not be resolved. A
realization of imaging with sub-wavelength resolution may enable
breakthrough in creation of future imaging devices at microwave,
sub-millimeter, terahertz, infrared and optical frequency ranges,
and can lead to significant progress of emerging technologies. For example, the
sub-wavelength imaging can be used in optical industry for
enlargement of the capacity of optical drives (DVD), in antenna
industry and in near field microscopy for improvement of near field
scanners, in medicine for magnetic resonance imaging (MRI) and
terahertz imaging, etc.

An idea to overcome the diffraction limit was first suggested by Sir
John Pendry in his seminal paper \cite{Pendrylens}. It was proposed
to use left-handed materials, isotropic media with both negative
permittivity and permeability \cite{Veselago}. A planar slab of such
a material provides an opportunity for imaging with sub-wavelength
resolution due to the effects of negative refraction and
amplification of evanescent waves. However, it is still problematic
to manufacture left-handed media since it requires to artificially
create negative permeability. Currently available designs of
left-handed media (at both microwave and optical frequency ranges)
are very lossy that restrict and even prevent their sub-wavelength
imaging applications. There is an alternative approach for
sub-wavelength imaging which does not involve either left-handed
media or effects of negative refraction and amplification of
evanescent waves. This new regime was recently suggested in the
paper \cite{canal} where it was named {\it canalization}. The idea
is that both propagating and evanescent harmonics of a source can be
transformed into the propagating waves inside a slab of certain
materials. Then, these propagating modes are capable of transporting
sub-wavelength images from one interface of the slab to another. The
source has to be placed very close to the front interface of the
slab in order to avoid deterioration of the sub-wavelength details
in the free space. A reflection from the slab can be minimized by
appropriate choice of its thickness. If the slab is tuned for
Fabry-Perot resonance then the reflection from its interface is
negligibly small for all angles of incidence including complex ones.
This allows to avoid harmful interactions between the source and the
slab which can disturb the image. An appropriate material for
operation in the canalization regime should have a flat isofrequency
contour, or in other words it should support waves traveling in a
certain direction with the fixed phase velocity for any transverse
wave vector. The materials which fulfil this requirement are
available at both microwave \cite{canal,Pekkaexp,SWIWM} and optical
\cite{layer} frequency ranges. The most interesting opportunity was
suggested in \cite{SWIWM} where the wire medium, a regular array of
parallel metallic wires \cite{WMPRB} was used. Originally, it was
expected that the transmission devices formed by the wire media
would operate only at microwave frequencies where the most of the
metals behave as ideal conductors. However, recently it has been
proposed to use the same structures for sub-wavelength imaging at
terahertz and infrared frequency bands where metals reveal their
plasma-like properties \cite{Marioplasm}.

In the present paper we study the performance of sub-wavelength
imaging by transmission devices formed by the wire medium at
microwave frequencies. The finite-sized planar slabs of wire medium
excited by sub-wavelength sources are modeled using the
finite-difference time-domain (FDTD) method. The wire medium is treated as
a bulky material other than an actual periodic array of wires
in \cite{SWIWM}.

\section{Spatial dispersion in the wire medium}
\begin{figure}[h]
\centering \epsfig{file=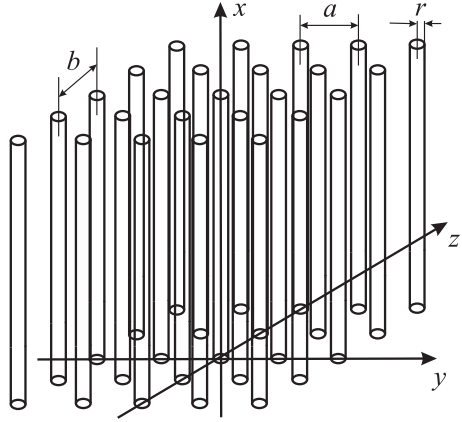, width=6cm} \caption{The wire
medium: a rectangular lattice of parallel ideally conducting thin
wires.} \label{geom}
\end{figure}

The wire medium is an artificial material formed by parallel perfect-
conducting wires arranged into a rectangular lattice (see Fig.
\ref{geom}).
This medium has been known as an artificial dielectric with
plasma-like properties at microwave frequencies for a long time
\cite{Rotmanps,Brown,pendryw}, but only recently it was shown that
the wire medium has strong spatial dispersion \cite{WMPRB}. In other
words, the wire medium is a non-local material and in the spectral
domain it can not be described using only frequency dependent
permittivity tensor. The permittivity tensor of the wire medium
depends also on the wave vector. In this paper we will use the
following expression for the permittivity tensor derived in
\cite{WMPRB}: \e \=\epsilon(\omega,\vec q)=\varepsilon(\omega,q_x)
\-x\-x+\-y\-y+\-z\-z, \qquad \varepsilon
(\omega,q_x)=1-\frac{k_p^2}{k^2-q_x^2}, \l{eff}\f where the $x$-axis
is oriented along wires, $q_x$ is the $x$-component of the wave
vector $\vec q$, $k=\omega/c$ is the wave number of the free space,
$c$ is the speed of light, and $k_p$ is the wave number
corresponding to the plasma frequency of the wire medium. The wave
number $k_p$ depends on the lattice periods $a$ and $b$, and on the
radius of wires $r$ \cite{WMJEWA}: \e
k_p^2=\frac{2\pi/(ab)}{\log\frac{\sqrt{ab}}{2\pi r}+F(a/b)}, \qquad
F(\xi)= -\frac{1}{2}\log \xi+ \sum\limits_{n=1}^{+\infty}
\left(\displaystyle \frac{\mbox{coth}(\pi n
\xi)-1}{n}\right)+\frac{\pi}{6}\xi \l{k0}. \f For the case of the
square grid ($a=b$), $F(1)=0.5275$ and the expression \r{k0} reduces
to \e k_p^2=\frac{2\pi/a^2}{\ln\frac{a}{2\pi r}+0.5275}. \l{k0sq}\f

The expression \r{eff} for the permittivity tensor of the wire
medium is valid if the wires are thin as compared to the lattice
periods (when the polarization across the wires is negligibly small
as compared to the longitudinal polarization) and if the lattice
periods are much smaller than the wavelength (when the wire medium
can be homogenized).

The wire medium supports three different types of modes, in contrast
to the usual uniaxial dielectrics which only support two types of
modes: ordinary and extraordinary:
\begin{itemize}
\item TE modes (ordinary modes, transverse electric field with respect to the wires), the waves which are polarized across the wires and
do not induce any currents along the wires. In the thin wire
approximation one can regard these as the modes which travel in
the free space and do not interact with the wires.

\item TM modes (extraordinary modes, transverse magnetic field with respect to the wires), the waves which correspond to nonzero currents
in the wires and nonzero electric field along the wires. At the
frequencies below the plasma frequency, these waves are evanescent.

\item TEM modes (transmission-line modes, transverse electric and magnetic field with respect to
the wires), the waves with nonzero
currents in the wires, but with zero electric field along the wires.
These modes travel with the speed of light along the wires ($q_x=\pm
k$) and can have any wave vector in the transverse direction.
Effectively, these waves correspond to the modes of a
multi-conductor transmission line formed by the wires, and enable us
to use wire medium for realization of the canalization regime.

\end{itemize}

The presence of TEM (transmission-line) mode provides an evidence of the
strong spatial dispersion in the wire media. Usually, effects of the
spatial dispersion can be observed in any periodical structure at the
frequencies corresponding to the spatial resonances when the
wavelength in the free space is comparable with the period of the
structure, as in usual photonic and electromagnetic crystals, or
when the wavelength in the crystal becomes comparable with the
period of the structure due to the resonant behavior of inclusions,
like in the case of resonant artificial dielectrics or photonic
crystals made from self-resonant inclusions. The wire medium is a
unique material where the spatial dispersion is observed at very low
frequencies and without any resonant effects.

\section{Spatially dispersive FDTD method for numerical modeling of the wire medium}
The FDTD method for numerical
solutions of Maxwell's equations was first proposed by Yee in 1966
\cite{Yee}. The method is based on an iterative process which allows
to obtain the electromagnetic fields throughout the computational
domain at a certain time step in terms of the fields at the previous
time steps using a set of updating equations \cite{Taflove}. The
typical discretization scheme involves forming a
dual-electric-magnetic field grid with electric and magnetic cells spatially
and temporally offset from each other. The main advantages of the FDTD
method are its simplicity, effectiveness and accuracy as well as the
capability of modeling frequency dispersive and anisotropic
materials. The FDTD method has been widely used for modeling of
various structures which include components formed by materials with
complex electromagnetic properties: frequency dispersive and
anisotropic dielectrics and magnetics, bi-anisotropic media,
ferrites, nonlinear materials, left-handed media and photonic crystals,
etc. The simulations of materials with complex inner geometries can
be performed either by modeling actual structures or through the
effective medium approach (if applicable). The second possibility
allows to dramatically decrease complexity of calculations and save
computational time.

In spite of the considerable number of complex electromagnetic media
modeled using the FDTD method, still there is a certain
incompleteness in consideration. Being focused on the frequency
dispersion effects, researchers have practically ignored the spatial
dispersion. According to our knowledge, so far no FDTD modeling of
spatially dispersive materials has been done. Probably, the lack of
attention to the spatial dispersion effects is caused by two
following reasons. The first reason is that it is widely assumed
that the spatial dispersion effects are usually quite weak and can
be neglected. However, this is not true: there is a number of
complex materials in which the spatial dispersion effects are strong
and play dominant roles in determination of their electromagnetic
properties. One example of such materials, the wire medium, is
considered in this paper. The second possible reason why the spatial
dispersion effects did not attract much attention of researchers is
a complexity in description of the spatial dispersion effects. The
analytical expressions which describe spatial dispersion for
particular materials are generally not available in contrast to the
numerous analytical models which describe the frequency dispersion
effects. The wire medium is an exception form this rule. The spatial
dispersion effects in this material can be described by simple
analytical formula \r{eff}.

In the present paper, the wire medium is modeled as a frequency and
spatially dispersive dielectric with permittivity tensor \r{eff}.
Both frequency and spatial dispersion effects are taken into account
in FDTD modeling using an auxiliary differential equation method. At
the spectral ($\omega$-$\vec q$) domain, the relation between the
electric flux density $\vec D$ and the electric field intensity
$\vec E$ in the wire medium is following: \e \vec D(\omega,\vec q)=
\varepsilon_0\=\varepsilon (\omega,\vec q) \vec E(\omega,\vec q),
\l{DE}\f where $\varepsilon_0$ is permittivity of the free space and
$\=\varepsilon (\omega,\vec q)$ is given by \r{eff}. The
substitution of \r{eff} into \r{DE} gives:
\begin{equation}
\left(k^{2}-q^{2}_{x}\right)D_{x}(\omega,\vec q)+\left(q^{2}_{x}
-k^{2}+k^{2}_{p}\right)\varepsilon_{0}E_{x}(\omega,\vec q)=0,
\label{eq_D_E2_x}
\end{equation}
\begin{equation}
D_y(\omega,\vec q)=\varepsilon_0E_y(\omega,\vec q),~~~~D_z(\omega,\vec q)=\varepsilon_0E_z(\omega,\vec q).
\label{eq_D_E2}
\end{equation}
Applying inverse Fourier transformation to Eqs. (\ref{eq_D_E2_x})
and (\ref{eq_D_E2}) we obtain the constitutive relations in the
time-space ($t$-$\vec r$) domain in the following form:
\begin{equation}
\left(\frac{\partial^{2}}{\partial
x^{2}}-\frac{1}{c^{2}}\frac{\partial^{2}} {\partial
t^{2}}\right)D_{x}(t,\vec
r)+\left(\frac{1}{c^{2}}\frac{\partial^{2}} {\partial
t^{2}}-\frac{\partial^{2}}{\partial x^{2}}+k^{2}_{p}\right)
\varepsilon_{0}E_{x}(t,\vec r)=0, \label{eq_D_E3_x}
\end{equation}
\begin{equation}
D_y(t,x)=\varepsilon_0E_y(t,\vec r),\qquad
D_z(t,x)=\varepsilon_0E_z(t,\vec r). \label{eq_D_E3}
\end{equation}

The FDTD simulation domain is represented by an equally spaced
three-dimensional grid with periods $\Delta_{x}$, $\Delta_{y}$ and
$\Delta_{z}$ along $x$-, $y$- and $z$-directions, respectively. The
time step is $\Delta_{t}$. For discretization of (\ref{eq_D_E3_x}),
we use the central finite difference operators in both time
$\delta^{2}_{t}$ and space $\delta^{2}_{x}$, defined as in
\cite{Hildebrand}:
\begin{eqnarray}
\frac{\partial^{2}}{\partial t^{2}}\rightarrow\frac{\delta^{2}_{t}}
{\Delta^{2}_{t}}, \qquad \delta^{2}_{t}F|^{n}_{m_{x},m_{y},m_{z}}&\equiv&F|^{n+1}_{m_{x},m_{y},m_{z}}-2F|^{n}_{m_{x},m_{y},m_{z}}+F|^{n-1}_{m_{x},m_{y},m_{z}};\nonumber\\
\frac{\partial^{2}}{\partial x^{2}}\rightarrow
\frac{\delta^{2}_{x}}{\Delta^{2}_{x}},\qquad
\delta^{2}_{x}F|^{n}_{m_{x},m_{y},m_{z}}&\equiv&F|^{n}_{m_{x}+1,m_{y},m_{z}}-2F|^{n}_{m_{x},m_{y},m_{z}}+F|^{n}_{m_{x}-1,m_{y},m_{z}};
   \label{eq_operators}
\end{eqnarray}
where $F$ represents $D_x$ or $E_x$; $m_{x},m_{y},m_{z}$ are indices
corresponding to a certain discretization point in the FDTD domain,
and $n$ is a number of the time steps. The discretization of Eq.
(\ref{eq_D_E3}) leads to the standard updating equations. The
discretized Eq. (\ref{eq_D_E3_x}) reads
\begin{equation}
\left(\frac{\delta^{2}_{x}}{\Delta^{2}_{x}}-\frac{1}{c^{2}}\frac{\delta^{2}_{t}}
{\Delta^{2}_{t}}\right)D_{x}+\left(\frac{1}{c^{2}}\frac{\delta^{2}_{t}}{\Delta^{2}_{t}}
-\frac{\delta^{2}_{x}}{\Delta^{2}_{x}}+k^{2}_{p}\right)\varepsilon_{0}E_{x}=0,
    \label{eq_D_E4_approx}
\end{equation}
and it can be rewritten in the open form as:
\begin{eqnarray}
\lefteqn{\!\biggl(\frac{D_{x}|^{n}_{m_{x}+1,m_{y},m_{z}}-2D_{x}|^{n}_{m_{x},m_{y},m_{z}}
+D_{x}|^{n}_{m_{x}-1,m_{y},m_{z}}}{\Delta^{2}_{x}}-\frac{1}{c^{2}}\frac{D_{x}|^{n+1}_{m_{x},m_{y},m_{z}}
-2D_{x}|^{n}_{m_{x},m_{y},m_{z}}+D_{x}|^{n-1}_{m_{x},m_{y},m_{z}}}{\Delta^{2}_{t}}\biggr)}\nonumber\\
&&\!\!\!\!\!\!\!\!\!\!\!\!\!\!\!\!+\varepsilon_{0}\biggl(\frac{1}{c^{2}}\frac{E_{x}|^{n+1}_{m_{x},m_{y},m_{z}}
-2E_{x}|^{n}_{m_{x},m_{y},m_{z}}+E_{x}|^{n-1}_{m_{x},m_{y},m_{z}}}{\Delta^{2}_{t}}
-\frac{E_{x}|^{n}_{m_{x}+1,m_{y},m_{z}}-2E_{x}|^{n}_{m_{x},m_{y},m_{z}}
+E_{x}|^{n}_{m_{x}-1,m_{y},m_{z}}}{\Delta^{2}_{x}}\nonumber\\
&&~~+k^{2}_{p}E_{x}|^{n}_{m_{x},m_{y},m_{z}}\biggr)=0.
    \label{eq_D_E4}
\end{eqnarray}
Therefore, the updating equation for $E_{x}$ in terms of $E_{x}$ and
$D_{x}$ at the previous time steps is obtained as follows:
\begin{eqnarray}
\lefteqn{E_{x}|^{n+1}_{m_{x},m_{y},m_{z}}=\frac{c^{2}\Delta^{2}_{t}}{\Delta^{2}_{x}}
\left(E_{x}|^{n}_{m_{x}+1,m_{y},m_{z}}+E_{x}|^{n}_{m_{x}-1,m_{y},m_{z}}\right)
+\left(2-\frac{2c^{2}\Delta^{2}_{t}}{\Delta^{2}_{x}}-c^{2}\Delta^{2}_{t}k^{2}_{p}\right)E_{x}|^{n}_{m_{x},m_{y},m_{z}}}\nonumber\\
&&~~~~~~~~~~~~~~-E_{x}|^{n-1}_{m_{x},m_{y},m_{z}}+\varepsilon^{-1}_{0}\biggl[D_{x}|^{n+1}_{m_{x},m_{y},m_{z}}
+D_{x}|^{n-1}_{m_{x},m_{y},m_{z}}
-2\left(1-\frac{c^{2}\Delta^{2}_{t}}{\Delta^{2}_{x}}\right)D_{x}|^{n}_{m_{x},m_{y},m_{z}}\nonumber\\
&&~~~~~~~~~~~~~~~~~~~~~~~~~~~~~~~~~~~~~~~~~~~~~~~-\frac{c^{2}\Delta^{2}_{t}}{\Delta^{2}_{x}}\left(D_{x}|^{n}_{m_{x}+1,m_{y},m_{z}}+D_{x}|^{n}_{m_{x}-1,m_{y},m_{z}}\right)
\biggr]. \label{update}
\end{eqnarray}
In the spatially dispersive FDTD modeling of the wire medium, the
calculations of \vec{D} from the magnetic field intensity \vec{H},
and \vec{H} from \vec{E} are performed using Yee's standard FDTD
equations \cite{Taflove}, while $E_{x}$ is calculated from $D_{x}$
using (\ref{update}) and $E_{y}=\varepsilon^{-1}_{0}D_{y}$,
$E_{z}=\varepsilon^{-1}_{0}D_{z}$. The particular details concerning
numerical aspects of proposed spatially dispersive FDTD method such
as stability analysis and numerical dispersion are available in
\cite{IEEEYan}.

The Eq. (\ref{update}) incorporates the terms corresponding to both
frequency and spatial dispersion. If the terms corresponding to the
spatial dispersion effects (which contain $\Delta_{x}$) are omitted
from Eq. (\ref{update}) then the rest of expression gives classical
updating equation for the Drude material with the collision
frequency equal to zero i.e. $\varepsilon(\omega)=1-k_p^2/k^2$
\cite{Prokopidis}:
\begin{eqnarray}
\lefteqn{E_{x}|^{n+1}_{m_{x},m_{y},m_{z}}=
\left(2-c^2\Delta^{2}_{t}k_p^2\right)E_{x}|^{n}_{m_{x},m_{y},m_{z}}-E_{x}|^{n-1}_{m_{x},m_{y},m_{z}}}\nonumber\\
&&~~~~~~~~~~~~~+\varepsilon_0^{-1}\left[D_{x}|^{n+1}_{m_{x},m_{y},m_{z}}-2D_{x}|^{n}_{m_{x},m_{y},m_{z}}
+D_{x}|^{n-1}_{m_{x},m_{y},m_{z}}\right].
\label{updatedrude}
\end{eqnarray}
In this paper we will use the updating Eq. (\ref{updatedrude}) for
modeling of uniaxial Drude material (frequency dispersive material
with no spatial dispersion) which can be treated as conventional and
incorrect description of the wire medium \cite{Rotmanps,Brown,pendryw},
in order to demonstrate the significance of taking into account the
spatial dispersion effects in the modeling of the wire medium.

We would like to note that in the present paper we consider only
so-called one-dimensional wire media, the lattices of parallel
ideally conducting wires oriented in one direction as in
Fig.~\ref{geom}, which can be described following \cite{WMPRB} by
permittivity tensor \r{eff} with one spatially dispersive component.
Actually, there are also two- and three-dimensional wire media
\cite{2d3dWM,Mariohomo1,Mariohomo2} consisting of two or three
orthogonal non-connected lattices of parallel ideally conducting
wires, which can be described by permittivity tensors with two or
three spatially dispersive components of the same form as in
\r{eff}, respectively. These two- and three-dimensional wire media
can be easily modeled using the spatially dispersive FDTD method as
introduced above: the updating equation (\ref{update}) has to be
applied in two or three directions, respectively.

\section{Excitation of the wire medium slab by a point magnetic source}
\begin{figure}[h]
\centering \epsfig{file=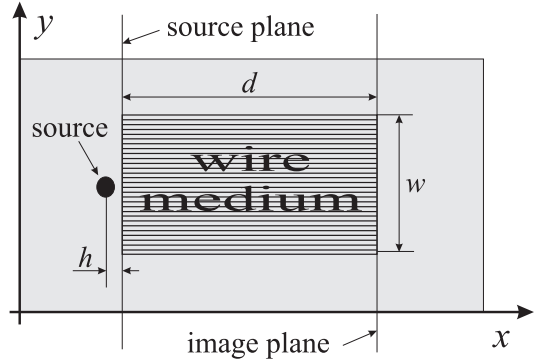, width=7cm} \caption{The layout
of the computation domain for two-dimensional FDTD simulations}
\label{domain}
\end{figure}

We have implemented the spatially dispersive FDTD method in two-dimensional
simulations in order to study the wave propagation through the wire
medium. The computation domain is infinite in $z$-direction and has
a rectangular shape in $x-y$ plane (see Fig. \ref{domain}). A
ten-cell Berenger's perfectly matched layer (PML) \cite{Berenger} is
used to truncate the computation domain. The FDTD cell size is
$\Delta_{x}=\Delta_{y}=\lambda/200$ and the time step according to
the Courant stability condition \cite{Taflove} is
$\Delta_{t}=\Delta_{x}/2c$.
The polarization is transverse magnetic (TM) with respect to the
orientation of the wires (or transverse electric (TE) with respect
to the computation domain): The electric field is in the $x-y$ plane
and the magnetic field is along $z$-axis. The wire medium has a
rectangular shape with dimensions $d\times w$ in $x-y$ plane (see
Fig. \ref{domain}). The wires are oriented along $x$-axis. The wave
number corresponding to the plasma frequency of the wire medium is
chosen to be four times larger than the wave number of the free
space ($k_{p}=4k$). The thickness of the slab $d$ has to be an
integer number of half-wavelengths in order to implement the
canalization regime \cite{canal,SWIWM}. A point magnetic source
radiating sinusoidal signal with free space wavelength $\lambda$ is placed at
the distance $d$ from the front interface of the wire medium slab.
As illustrated in Fig. \ref{domain}, the front interface of the
imaging device is treated as the source plane, and the back
interface as the image plane, respectively.

\subsection{Transient wave propagation}
\begin{figure}[h]
\centering \epsfig{file=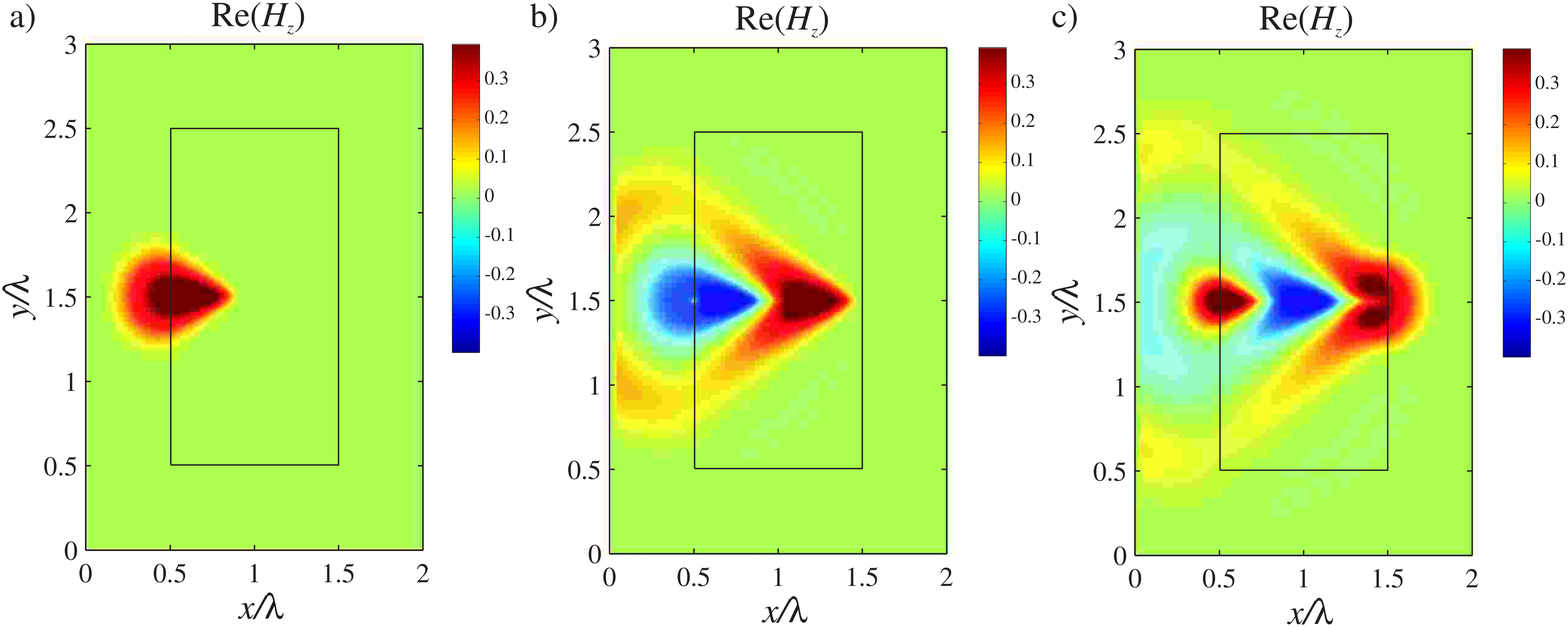, width=13cm}
\caption{Animation: Transient propagation of the wave excited by a
point sinusoidal magnetic source located at the $\lambda/200$
distance from the front interface of $1\lambda\times 2\lambda$ slab
of the wire medium. Figure: Snapshots of magnetic field $H_z$ at the
different time steps: (a) $t=175\Delta_{t}$, (b) $t=400\Delta_{t}$,
(c) $t=530\Delta_{t}$.\vspace{-3mm}} \label{wave_front}
\end{figure}

In the first simulation, see Fig. \ref{wave_front}, we have placed
the point source in the close vicinity from the front interface of
the wire medium slab ($h=\lambda/200$). Such a geometry allows us to
illustrate the fact that the waves in the wire medium travel in the
fixed direction (along the wires) with the speed of light, in
contrast to the free space case where the waves travel with the
speed of light in arbitrary direction. The point source creates a
cylindrical wavefront in the free space (see  Fig. \ref{wave_front})
since the waves are allowed to travel in all directions with the
same speed. However, as soon as the cylindrical wave enters the wire
medium, the form of wavefront changes dramatically (see Fig.
\ref{wave_front}): the wavefront becomes conical. The waves travel
with speed of light along the surface of the wire medium in the free
space and excites transmission line modes of the wire medium which
travel in the orthogonal direction (along the wires) with the same
speed, that is why the form of the wavefront happens to be conical.
Note that the front part of the cone contains sub-wavelength
information of the source and as soon as the cone reaches the back
interface of the slab one can state that the sub-wavelength image is
formed. Thus, we conclude that the formation of images happens with
the speed of light and all spatial harmonics from the spectrum of
the source reaches the image plane at the same time. This fact
illustrates the advantage of the canalization principle of
sub-wavelength imaging over the regime based on negative refraction
and amplification of evanescent waves, where the pumping of the
growing evanescent waves takes very long and the sub-wavelength
information corresponding to the different spatial harmonics appears
at the image plane with significant delay.

\subsection{Power flow}
\begin{figure}[b]
\vspace{-7mm} \centering \epsfig{file=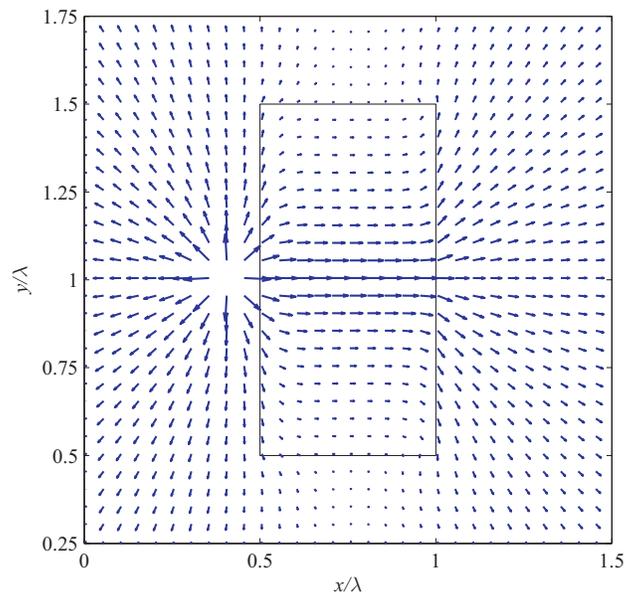, width=11.5cm}
\vspace{-3mm} \caption{Power flow diagram in the steady state for
the $0.5\lambda\times 1\lambda$ slab of the wire medium excited by a
point magnetic source located at the $\lambda/10$ distance from the
front interface.} \label{power_flow}
\end{figure}

In the second simulation, we have placed a point magnetic source at
the $\lambda/10$ distance from the $0.5\lambda\times 1\lambda$ slab
of wire medium. The power flow diagram in the steady-state regime
for this case is presented in Fig. \ref{power_flow}.
One can see that in the close vicinity of the interfaces the power
flow changes direction because of the evanescent extraordinary modes
of the wire medium and inside the wire medium the energy transfer
happens only along direction of wires with the help of transmission
line modes. Also, it is noteworthy that no harmful diffractions from
the corners of the slab are observed. This can be explained by the
fact that the waves in the wire medium travel and transfer energy
only along $x$-axis and no waves travel along $y$-axis. That is why
the interfaces in $y$-direction does not reflect any waves and no
diffractions from the corners are visible. Due to the absence of
diffraction effects it is not mandatory to have the transverse size
$w$ of the slab to be significantly larger than the wavelength in
order to provide functionality of the transmission device, in
contrast to the case of conventional lenses. The transverse
dimensions of the transmission device can be practically arbitrary.
For example, in the present paper we use one- and two-wavelength
wide slabs which demonstrate good sub-wavelength imaging
performance.

\subsection{Distributions of electric and magnetic fields}
\begin{figure}[h]
\centering \epsfig{file=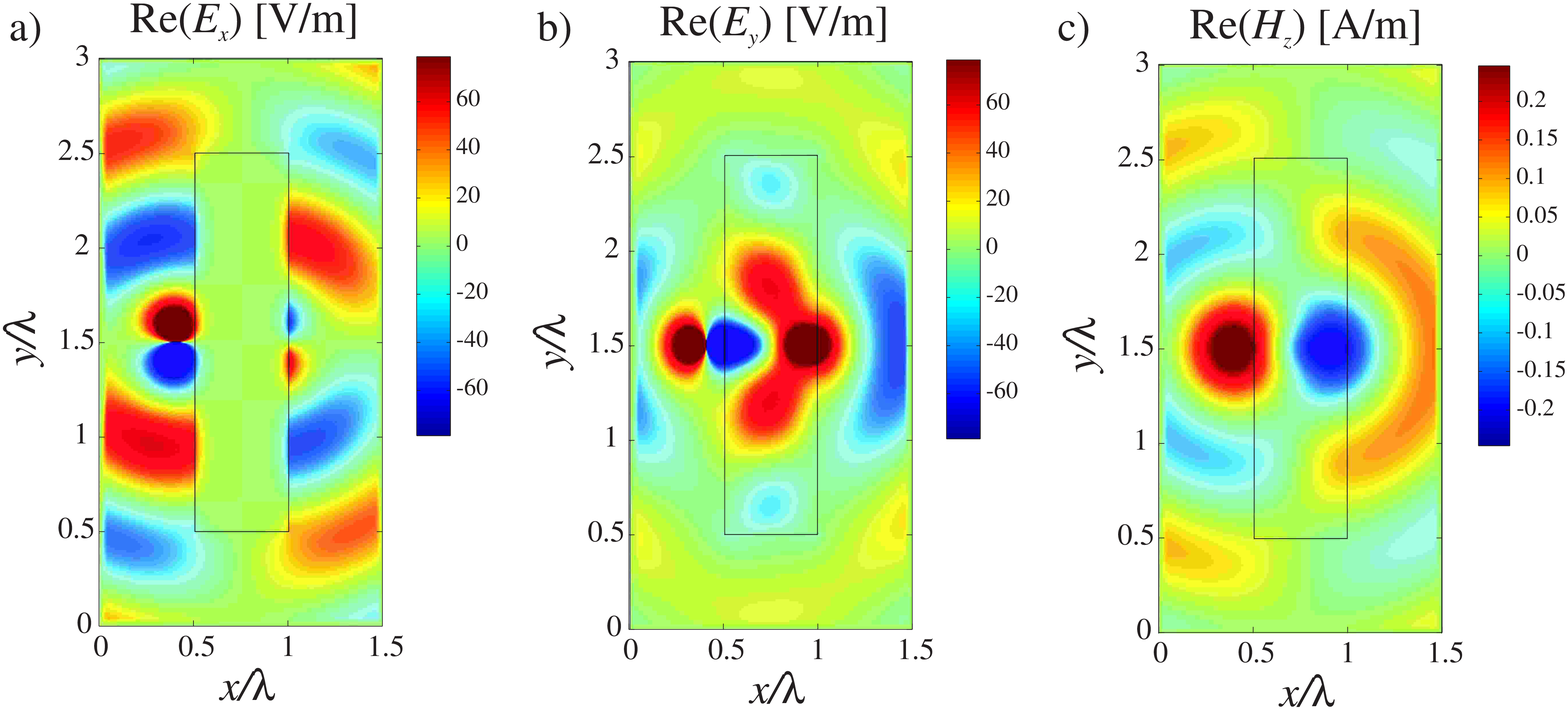, width=13.2cm} \caption{Animation:
Distributions of electric and magnetic fields for a
$0.5\lambda\times 2\lambda$ slab of the wire medium excited by a
point source located at $\lambda/10$ distance from the front
interface.} \label{wm1_r}
\end{figure}

\begin{figure}[h]
\centering \epsfig{file=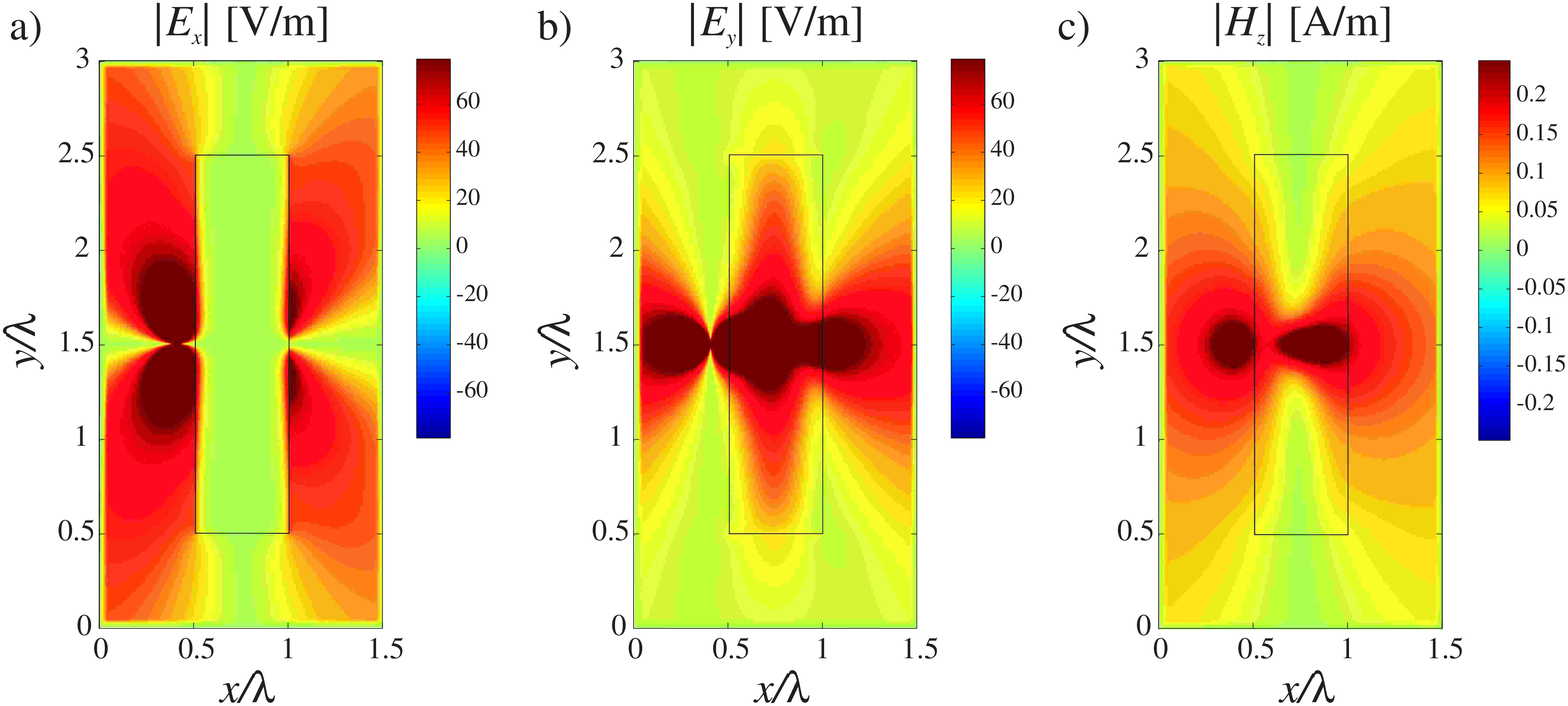, width=13.2cm} \caption{Absolute
values of the fields plotted in  Fig. \ref{wm1_r}} \label{wm1_a}
\end{figure}

In the next simulation, see Fig. \ref{wm1_r}, we have kept
$\lambda/2$ thickness of the slab and $\lambda/10$ distance between
source and the wire medium slab, but increased transverse size of
the slab up to $2\lambda$. Fig. \ref{wm1_r} shows the distributions
of electric and magnetic fields after the steady state is reached.
The absolute values of these fields are presented in Fig.
\ref{wm1_a}.
One can see from Fig. \ref{wm1_r}.a that the non-zero $x$-component
of electrical field is present inside the wire medium slab only in
the close vicinity of the interfaces. This can be easily explained
since only the extraordinary modes of the wire medium have non-zero
electric field along the wires, but these modes in the present case
are evanescent and decay with distance. That is why the
$x$-component of electrical field vanishes in the center of the
slab. Inside the wire medium slab, only transmission line modes are
present and thus, the electric and magnetic fields have only $y$-
and $z$-components, respectively, see Figs. \ref{wm1_r}.b and
\ref{wm1_r}.c. In accordance to the canalization principle, the
transmission line modes deliver images from the front interface to
the back one. This is why the absolute values of the fields are the
same at the front and back interfaces, see Fig. \ref{wm1_a}.
However, the fields in the image plane appear in out of phase with
respect to the fields in the source plane since the thickness of the
slab is $\lambda/2$.

\subsection{Effect of the slab thickness}
The images produced by transmission devices operating in the
canalization regime exactly repeat the source distributions if the
thickness of the structure is equal to even number of
half-wavelengths and appear out of phase if the thickness is
equal to odd number of half-wavelengths \cite{canal}. In order to
demonstrate the case when the image appears in phase with the source
we have chosen a wire medium slab with the dimension of
$2\lambda\times1\lambda$. Other parameters are the same as in the
previous case.
\begin{figure}[b]
\vspace{-2mm} \centering \epsfig{file=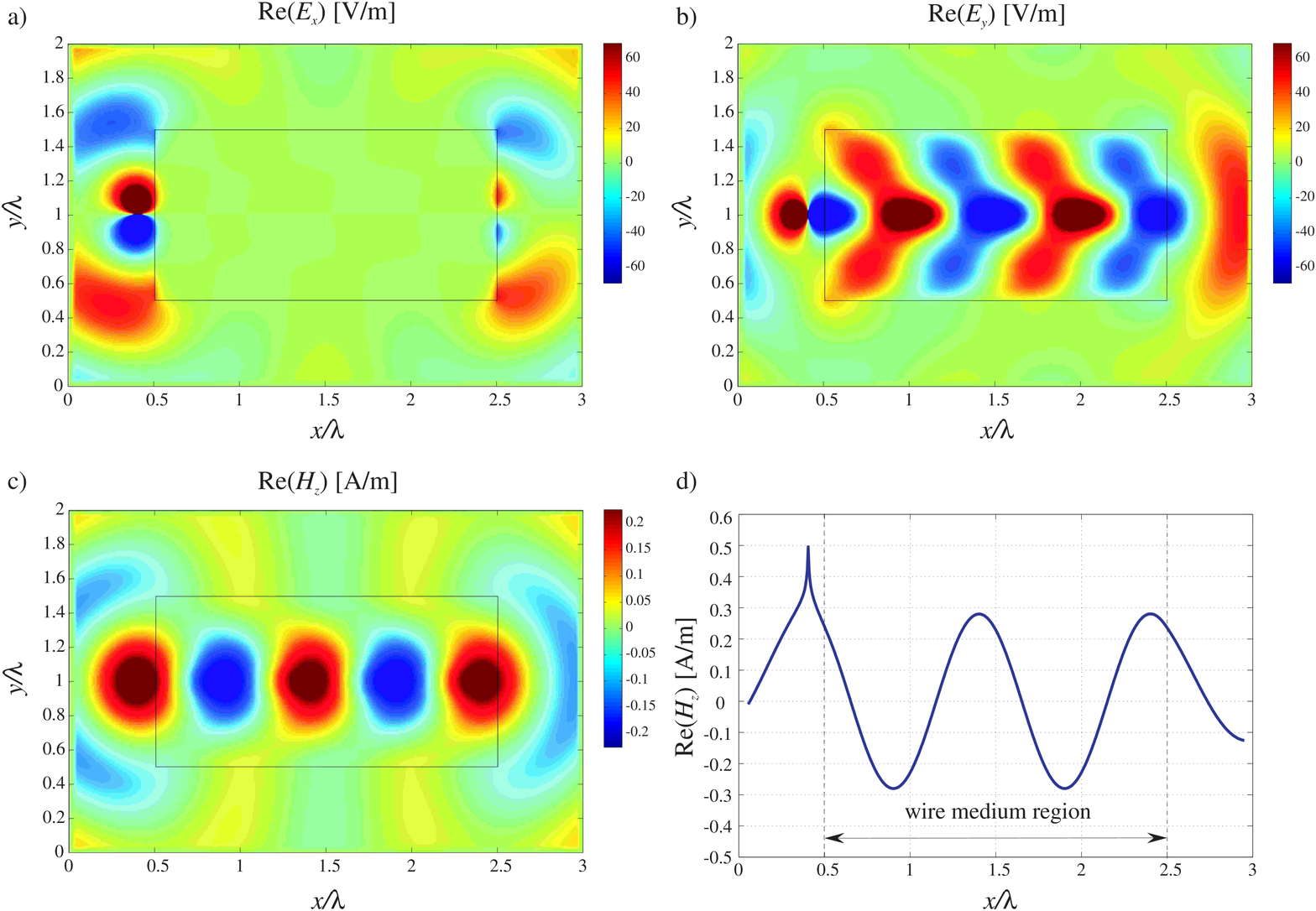, width=13.2cm}
\caption{Distributions of electric and magnetic fields for a
$2\lambda\times \lambda$ slab of the wire medium excited by a point
source located at $\lambda/10$ distance from the front interface.}
\label{wm4}
\end{figure}
Figs. \ref{wm4}.a, b and c show the distributions of electric and
magnetic field in the simulation domain for this case and it is
clearly seen that the fields at the back interface of the slab are
the same as at the front interface. Additionally, the distribution
of magnetic field in the plane $y=\lambda$ is plotted in Fig.
\ref{wm4}.d in order to demonstrate that the field inside of the
wire medium has harmonic dependence with the period $\lambda$ along
the direction of wires. Using this illustration and the fact of
magnetic field continuity at the interface between free space and
the wire medium, it becomes possible to explain the canalization
principle for sub-wavelength imaging: The field at the back
interface of the slab repeats the distribution at the front
interface because the total thickness of the slab is equal to
integer numbers of half-wavelength and at such a distance any
harmonic dependence with period $\lambda$ repeats its initial
absolute value.

\section{Effects of spatial dispersion in modeling of the wire medium}
\begin{figure}[h]
\centering \epsfig{file=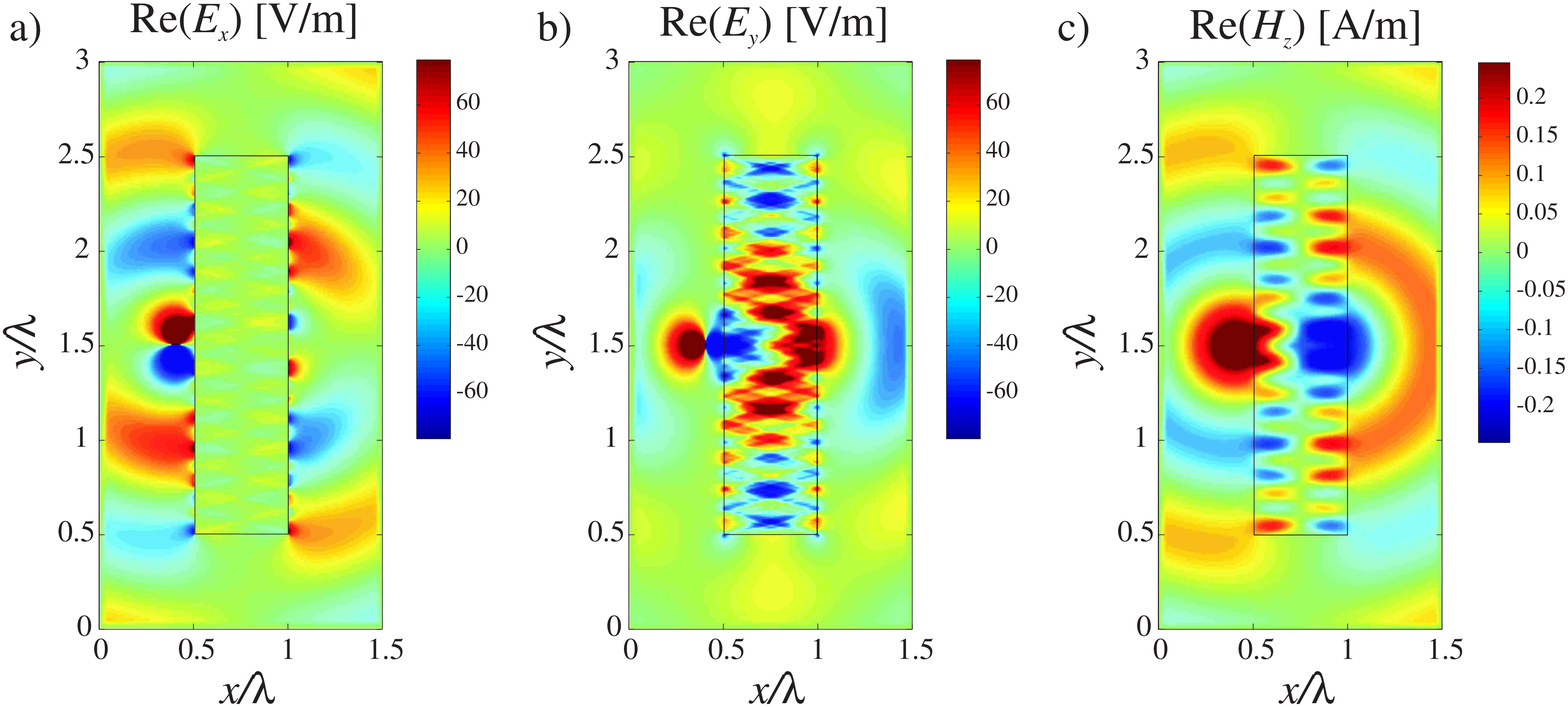, width=13.2cm}
\caption{Distributions of electric and magnetic field in the steady
state for a $0.5\lambda\times 2\lambda$ slab of the uniaxial Drude
material \r{epsl} with $k_{p}/k=4$ excited by a point source located
at $\lambda/10$ distance from the front interface.} \label{wm2_1}
\end{figure}

The wire medium is a unique artificial material which possesses
strong spatial dispersion effects. It is quite hard to compare
properties of this material with media without spatial dispersion,
however it is possible to reveal certain similarities and
differences. For example, if the spatial dispersion effects are
neglected in \r{eff} then one deals with uniaxial Drude material
with permittivity tensor \e
\=\varepsilon=\varepsilon(\omega)\-x\-x+\-y\-y+\-z\-z, \qquad
\varepsilon(\omega)=1-\frac{k_p^2}{k^2}.\l{epsl}\f Such a model can
be treated as an old and incorrect description of the wire medium
\cite{Rotmanps,Brown,pendryw}. We have replaced the wire medium in
simulations corresponding to Fig. \ref{wm1_r} by a local uniaxial
Drude material with permittivity tensor \r{epsl}. All other
parameters of the structure ($k_p=4k$, $h=\lambda/10$, $w=2\lambda$,
$d=\lambda/2$, see Fig. \ref{domain}) were kept unchanged. The FDTD
simulation was performed using the updating equation
(\ref{updatedrude}). The distributions of the fields in the
steady-state regime are presented in  Fig. \ref{wm2_1}.
One can readily notice strong differences between this figure and
Fig. \ref{wm1_r} corresponding to the case of the wire medium: The
waves inside of the uniaxial Drude material travel not precisely
along the anisotropy axis, and thus, they suffer numerous
reflections from the edges and corners of the slab. The interference
pattern caused by the multiple-reflected waves distorts field
distributions in both source and image planes, see Fig.
\ref{Hz_cut_abs}.b. This example shows that it is extremely
important to take into account the spatial dispersion in modeling of
the wire media.

\begin{figure}[h]
\centering \epsfig{file=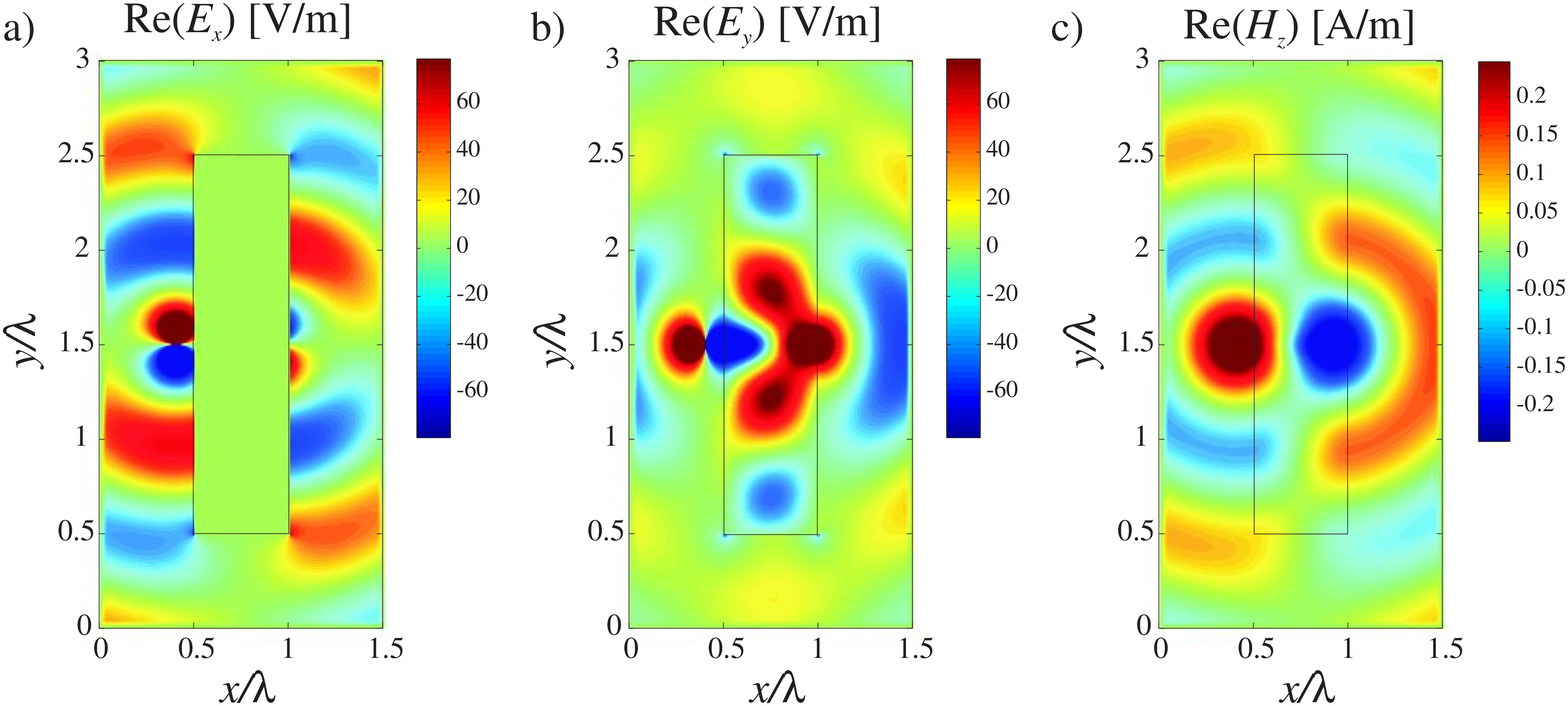, width=13.2cm}
\caption{Distributions of electric and magnetic fields in the steady
state for a slab of the uniaxial material with infinite permittivity
along anisotropy axis \r{epsinf} excited by a point source located
at $\lambda/10$ distance from the front interface.} \label{wm2_2}
\end{figure}

\begin{figure}[h]
\centering \epsfig{file=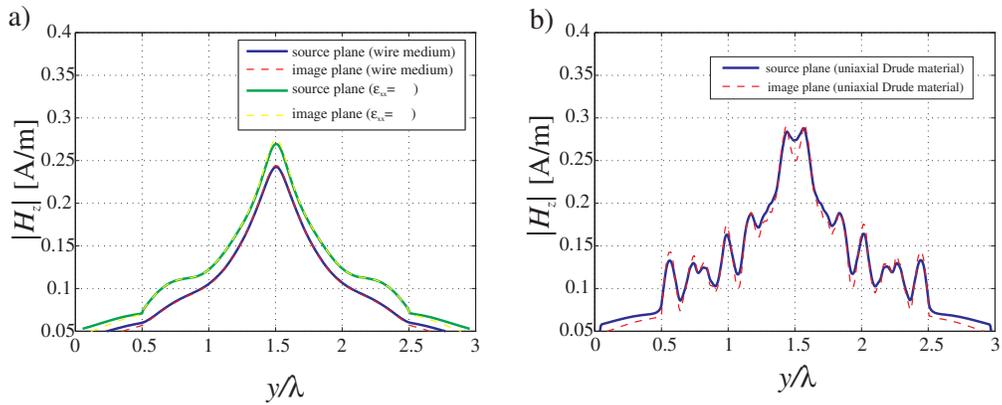, width=13.2cm}
\caption{Absolute values of magnetic field in the source and image
planes of the transmission devices formed by three different
materials: a) wire medium (Fig. \ref{wm1_a}), uniaxial material with
infinite permittivity along anisotropy axis (Fig.~\ref{wm2_2}); b)
uniaxial Drude material (Fig.~\ref{wm2_1}).} \label{Hz_cut_abs}
\end{figure}

Actually, the wire medium behaves similar to the uniaxial material
with permittivity tensor \e
\=\epsilon=\infty\-x\-x+\-y\-y+\-z\-z.\l{epsinf}\f This can be
explained by the fact that the component $\epsilon(\omega,q_x)$ of
the permittivity tensor of wire medium \r{eff} happens to be
infinite for transmission line modes with $q_x=\pm k$. We have
performed FDTD simulation for the structure as in Fig. \ref{wm1_r},
but instead of wire medium we used the uniaxial material with
infinite permittivity along the anisotropy axis \r{epsinf}. The
updating equation for $E_x$ component inside such a material is very
simple: $E_x|^{n+1}_{m_{x},m_{y},m_{z}}=0$. The results of the
simulation are presented in Fig. \ref{wm2_2}.

It first appears that Figs. \ref{wm2_2} and \ref{wm1_r} seem to be
identical. However, more precise comparison immediately reveals
significant differences: in the case of the uniaxial material with
infinite permittivity along anisotropy axis, $x$-component of
electric field is zero everywhere inside this material, but in the
case of wire medium there are regions near the front and back
interfaces of the slab where this component is non-zero. In other
words, the model \r{epsinf} allows to describe transmission line
modes of the wire medium, but it does not take into account
extraordinary modes of the wire medium. The presence of
extraordinary modes in the wire medium makes distributions of the
fields in the source and image plane slightly different from those
in the case of the uniaxial material with infinite permittivity
along the anisotropy axis, see Fig. \ref{Hz_cut_abs}.a.

The distributions of magnetic field in the source and image planes
for the simulations considered in this section are plotted in Fig.
\ref{Hz_cut_abs} for comparison. The slabs of wire medium and
uniaxial material with infinite permittivity along the anisotropy
axis demonstrate nearly perfect imaging: The distributions of the
field in the source and image planes are practically identical. The
difference between the distributions can be explained by the fact
that high-order spatial harmonics experience non-zero reflections
from the wire medium slab as it is shown in \cite{resolWM}. In the
case of local uniaxial Drude material, see Fig. \ref{Hz_cut_abs},
these reflections are significantly higher, and strong ripples
caused by interference of the waves reflected from the edges and
corners of the slab encumber proper imaging operation of the device.

\section{Sub-wavelength imaging by the wire medium slabs}
\begin{figure}[h]
\centering \epsfig{file=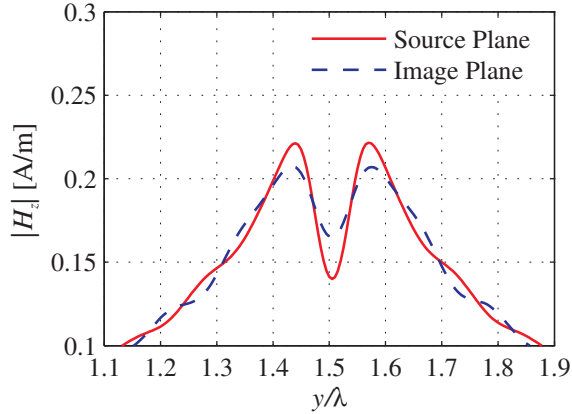, width=8.8cm}
\caption{Absolute values of magnetic field at the source and image
planes for a $0.5\lambda\times2\lambda$ slab of the wire medium
excited by three equally spaced magnetic sources with the phase
differences equal to $180^{\circ}$ located at $\lambda/20$ distance
from the front interface.} \label{Hz_cut_abs_3sources}
\end{figure}

In previous sections we have shown good imaging performance of the
wire medium slabs. However, we were not able to demonstrate the
sub-wavelength imaging capability of such devices since the point
source excitations were used in simulations. In order to investigate
such capability, we have performed additional simulations with more
complicated (sub-wavelength) sources. We have placed three equally
spaced magnetic point sources at $\lambda/20$ distance from the
front interface of a $0.5\lambda\times 2\lambda$ slab of wire
medium. The distance between sources is $\lambda/20$ and the central
source is excited in out of phase with respect to the neighboring
sources. The proposed three-source configuration creates
distribution with two strong maxima at the front interface of the
wire medium slab and the distance between these maxima is about
$\lambda/10$, see Fig. \ref{Hz_cut_abs_3sources}.
\begin{figure}[h]
\centering \epsfig{file=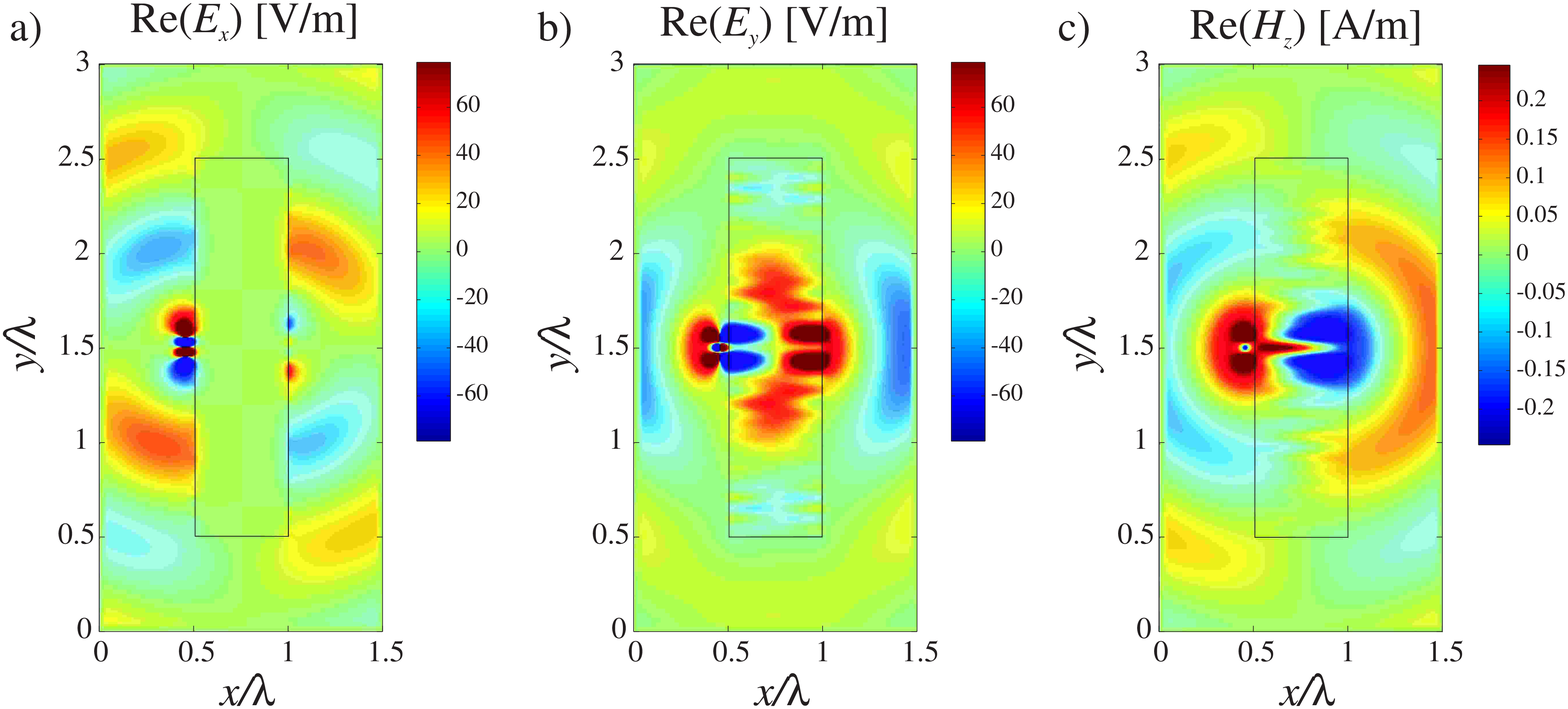, width=13.2cm} \caption{Animation:
Distributions of electric and magnetic fields in the steady state
for a $0.5\lambda\times2\lambda$ slab of the wire medium excited by
three equally spaced magnetic sources with the phase differences
equal to $180^{\circ}$ located at $\lambda/20$ distance from the
front interface.} \label{wm3}
\end{figure}
\begin{figure}[h]
\centering \epsfig{file=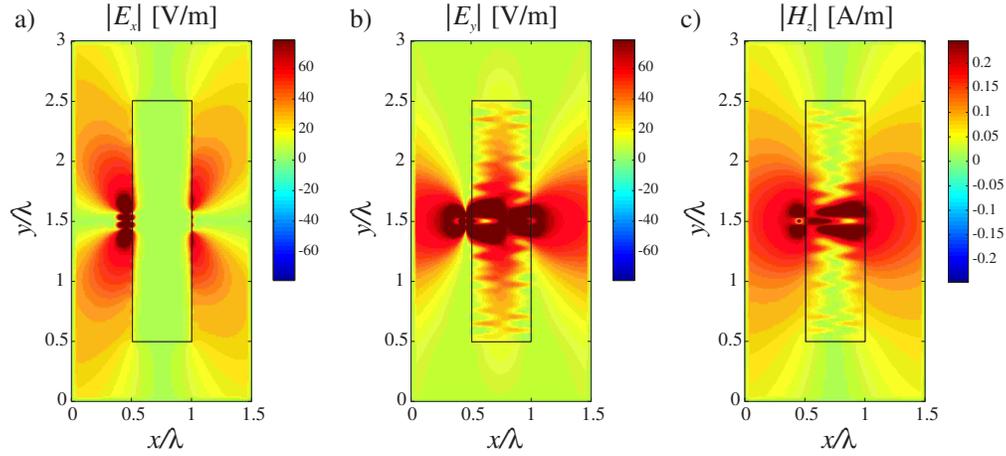, width=13.2cm} \caption{Absolute
values of the fields plotted in Fig. \ref{wm3}.} \vspace{-4mm}
\label{wm3_2}
\end{figure}
Results of the spatially dispersive FDTD simulation for excitation
of the wire medium slab by the proposed complicated source are
presented in Fig. \ref{wm3}. The size of the computation domain,
ratio between plasma frequency and operating frequency, FDTD cell
size and time step remain the same as in previous sections. The
distribution of magnetic field in the image plane is presented in
Fig. \ref{Hz_cut_abs_3sources}. The two maxima located at
$\lambda/10$ distance from each other are clearly resolved by the
device, that confirms the sub-wavelength imaging capability of the
wire medium lenses. For more details about resolution of such
imaging devices see theoretical investigations in \cite{resolWM}.

\section{Conclusion}
The spatially dispersive FDTD method has been developed for
efficient modeling of the wave propagation in the wire medium using
the effective medium approach. The auxiliary differential equation
method is used in order to take into account both the spatial and
frequency dispersion of the wire medium. The flat sub-wavelength
lenses formed by the wire medium are chosen for the validation of
developed spatially dispersive FDTD formulations. Numerical
simulations verify the sub-wavelength imaging capability of these
structures. The results confirm that the slabs of the wire medium
operate in the canalization regime as transmission devices and
demonstrate that this regime is not sensitive to the transverse
dimensions of these structures.

\section*{Acknowledgements}
P. A. Belov would like to thank Prof. Sergei Tretyakov for initial
suggestions on modeling the wire medium using the FDTD method.

\end{document}